


\font\twelverm=cmr10 scaled 1200    \font\twelvei=cmmi10 scaled 1200
\font\twelvesy=cmsy10 scaled 1200   \font\twelveex=cmex10 scaled 1200
\font\twelvebf=cmbx10 scaled 1200   \font\twelvesl=cmsl10 scaled 1200
\font\twelvett=cmtt10 scaled 1200   \font\twelveit=cmti10 scaled 1200
\font\twelvesc=cmcsc10 scaled 1200  
\skewchar\twelvei='177   \skewchar\twelvesy='60


\def\twelvepoint{\normalbaselineskip=12.4pt plus 0.1pt minus 0.1pt
  \abovedisplayskip 12.4pt plus 3pt minus 9pt
  \belowdisplayskip 12.4pt plus 3pt minus 9pt
  \abovedisplayshortskip 0pt plus 3pt
  \belowdisplayshortskip 7.2pt plus 3pt minus 4pt
  \smallskipamount=3.6pt plus1.2pt minus1.2pt
  \medskipamount=7.2pt plus2.4pt minus2.4pt
  \bigskipamount=14.4pt plus4.8pt minus4.8pt
  \def\rm{\fam0\twelverm}          \def\it{\fam\itfam\twelveit}%
  \def\sl{\fam\slfam\twelvesl}     \def\bf{\fam\bffam\twelvebf}%
  \def\mit{\fam 1}                 \def\cal{\fam 2}%
  \def\sc{\twelvesc}               \def\tt{\twelvett}
  \def\sf{\twelvesf}
  \textfont0=\twelverm   \scriptfont0=\tenrm   \scriptscriptfont0=\sevenrm
  \textfont1=\twelvei    \scriptfont1=\teni    \scriptscriptfont1=\seveni
  \textfont2=\twelvesy   \scriptfont2=\tensy   \scriptscriptfont2=\sevensy
  \textfont3=\twelveex   \scriptfont3=\twelveex  \scriptscriptfont3=\twelveex
  \textfont\itfam=\twelveit
  \textfont\slfam=\twelvesl
  \textfont\bffam=\twelvebf \scriptfont\bffam=\tenbf
  \scriptscriptfont\bffam=\sevenbf
  \normalbaselines\rm}



\def\beginlinemode{\endmode
  \begingroup\parskip=0pt \obeylines\def\\{\par}\def\endmode{\par\endgroup}}
\def\beginparmode{\endmode
  \begingroup \def\endmode{\par\endgroup}}
\let\endmode=\par
{\obeylines\gdef\
{}}
\def\singlespace{\baselineskip=\normalbaselineskip}

\def\oneandahalfspace{\baselineskip=\normalbaselineskip
  \multiply\baselineskip by 3 \divide\baselineskip by 2}
\def\doublespace{\baselineskip=\normalbaselineskip \multiply\baselineskip by 2}

\newcount\firstpageno
\firstpageno=2
\footline={\ifnum\pageno<\firstpageno{\hfil}\else{\hfil\twelverm\folio\hfil}\fi}
\def\toppageno{\global\footline={\hfil}\global\headline
  ={\ifnum\pageno<\firstpageno{\hfil}\else{\hfil\twelverm\folio\hfil}\fi}}
\let\rawfootnote=\footnote              
\def\footnote#1#2{{\rm\singlespace\parindent=0pt\parskip=0pt
  \rawfootnote{#1}{#2\hfill\vrule height 0pt depth 6pt width 0pt}}}
\def\raggedcenter{\leftskip=4em plus 12em \rightskip=\leftskip
  \parindent=0pt \parfillskip=0pt \spaceskip=.3333em \xspaceskip=.5em
  \pretolerance=9999 \tolerance=9999
  \hyphenpenalty=9999 \exhyphenpenalty=9999 }
\def\dateline{\rightline{\ifcase\month\or
  January\or February\or March\or April\or May\or June\or
  July\or August\or September\or October\or November\or December\fi
  \space\number\year}}
\def\received{\vskip 3pt plus 0.2fill
 \centerline{\sl (Received\space\ifcase\month\or
  January\or February\or March\or April\or May\or June\or
  July\or August\or September\or October\or November\or December\fi
  \qquad, \number\year)}}


\hsize=6.5truein
\vsize=8.5truein  
\voffset=0.0truein
\parskip=\medskipamount
\def\\{\cr}
\twelvepoint            
\doublespace            
\overfullrule=0pt       




\def\title                      
  {\null\vskip 3pt plus 0.2fill
   \beginlinemode \doublespace \raggedcenter \bf}

\def\author                     
  {\vskip 3pt plus 0.2fill \beginlinemode
   \singlespace \raggedcenter\sc}

\def\affil                      
  {\vskip 3pt plus 0.1fill \beginlinemode
   \oneandahalfspace \raggedcenter \sl}

\def\abstract                   
  {\vskip 3pt plus 0.3fill \beginparmode
   \singlespace ABSTRACT: }

\def\endtopmatter               
  {\endpage                     
   \body}

\def\body                       
  {\beginparmode}               

\def\head#1{                    
  \goodbreak\vskip 0.5truein    
  {\immediate\write16{#1}
   \raggedcenter \uppercase{#1}\par}
   \nobreak\vskip 0.25truein\nobreak}

\def\subhead#1{                 
  \vskip 0.25truein             
  {\raggedcenter {#1} \par}
   \nobreak\vskip 0.25truein\nobreak}

\def\beginitems{
\par\medskip\bgroup\def\i##1 {\item{##1}}\def\ii##1 {\itemitem{##1}}
\leftskip=36pt\parskip=0pt}
\def\enditems{\par\egroup}

\def\beneathrel#1\under#2{\mathrel{\mathop{#2}\limits_{#1}}}

\def\refto#1{$^{#1}$}           

\def\references                 
  {\head{{\bf References}}            
   \beginparmode
   \frenchspacing \parindent=0pt \leftskip=1truecm
   \parskip=8pt plus 3pt \everypar{\hangindent=\parindent}}

\gdef\refis#1{\item{#1.\ }}                     

\gdef\journal#1, #2, #3, 1#4#5#6{               
    {\sl #1~}{\bf #2}, #3 (1#4#5#6)}            

\gdef\refa#1, #2, #3, #4, 1#5#6#7.{\noindent#1, #2 {\bf #3}, #4 (1#5#6#7).\rm}

\gdef\refb#1, #2, #3, #4, 1#5#6#7.{\noindent#1 (1#5#6#7), #2 {\bf #3}, #4.\rm}

\def\pr{\journal Phys.Rev., }

\def\jmp{\journal J.Math.Phys., }

\def\endreferences{\body}

\def\figurecaptions             
  {\endpage
   \beginparmode
   \head{Figure Captions}
}

\def\endpage                    
  {\vfill\eject}

\def\endpaper                   
  {\endmode\vfill\supereject}


\def\heading                            
  {\vskip 0.5truein plus 0.1truein      
   \beginparmode \def\\{\par} \parskip=0pt \singlespace \raggedcenter}

\def\subheading                         
  {\vskip 0.25truein plus 0.1truein     
   \beginlinemode \singlespace \parskip=0pt \def\\{\par}\raggedcenter}

\def\tag#1$${\eqno(#1)$$}

\def\align#1$${\eqalign{#1}$$}

\def\aligntag#1$${\gdef\tag##1\\{&(##1)\cr}\eqalignno{#1\\}$$
  \gdef\tag##1$${\eqno(##1)$$}}

\def\overset #1\to#2{{\mathop{#2}\limits^{#1}}}
\def\underset#1\to#2{{\let\next=#1\mathpalette\undersetpalette#2}}
\def\undersetpalette#1#2{\vtop{\baselineskip0pt
\ialign{$\mathsurround=0pt #1\hfil##\hfil$\crcr#2\crcr\next\crcr}}}


\def\ref#1{Ref.~#1}                     
\def\Ref#1{Ref.~#1}                     
\def\[#1]{[\cite{#1}]}
\def\cite#1{{#1}}
\def\(#1){(\call{#1})}
\def\call#1{{#1}}
\def\taghead#1{}
\def\frac#1#2{{#1 \over #2}}
\def\half{{\frac 12}}

\def\12{{1\over2}}

\def\sla{\raise.15ex\hbox{$/$}\kern-.57em}
\def\leaderfill{\leaders\hbox to 1em{\hss.\hss}\hfill}
\def\twiddle{\lower.9ex\rlap{$\kern-.1em\scriptstyle\sim$}}
\def\bigtwiddle{\lower1.ex\rlap{$\sim$}}
\def\gtwid{\mathrel{\raise.3ex\hbox{$>$\kern-.75em\lower1ex\hbox{$\sim$}}}}
\def\ltwid{\mathrel{\raise.3ex\hbox{$<$\kern-.75em\lower1ex\hbox{$\sim$}}}}
\def\square{\kern1pt\vbox{\hrule height 1.2pt\hbox{\vrule width 1.2pt\hskip 3pt
   \vbox{\vskip 6pt}\hskip 3pt\vrule width 0.6pt}\hrule height 0.6pt}\kern1pt}
\def\tdot#1{\mathord{\mathop{#1}\limits^{\kern2pt\ldots}}}

\def\pmb#1{\setbox0=\hbox{#1}%
  \kern-.025em\copy0\kern-\wd0
  \kern  .05em\copy0\kern-\wd0
  \kern-.025em\raise.0433em\box0 }

\def\a{{\alpha}}

\def\D{{\cal D}}
\def\s{\sigma}
\def\half{{1 \over 2}}
\def\ra{{\rangle}}
\def\la{{\langle}}

\def\ih{{i \over \hbar}}

\def\E{{\cal E}}
\def\au{\underline \alpha}
\def\p{{\bf p}}

\def\x{{\bf x}}

\def\ria{{\rightarrow}}
\def\Tr{{\rm Tr}}

\catcode`@=11
\newcount\r@fcount \r@fcount=0
\newcount\r@fcurr
\immediate\newwrite\reffile
\newif\ifr@ffile\r@ffilefalse
\def\w@rnwrite#1{\ifr@ffile\immediate\write\reffile{#1}\fi\message{#1}}

\def\writer@f#1>>{}
\def\referencefile{
  \r@ffiletrue\immediate\openout\reffile=\jobname.ref%
  \def\writer@f##1>>{\ifr@ffile\immediate\write\reffile%
    {\noexpand\refis{##1} = \csname r@fnum##1\endcsname = %
     \expandafter\expandafter\expandafter\strip@t\expandafter%
     \meaning\csname r@ftext\csname r@fnum##1\endcsname\endcsname}\fi}%
  \def\strip@t##1>>{}}

\def\citeall#1{\xdef#1##1{#1{\noexpand\cite{##1}}}}
\def\cite#1{\each@rg\citer@nge{#1}} 

\def\each@rg#1#2{{\let\thecsname=#1\expandafter\first@rg#2,\end,}}
\def\first@rg#1,{\thecsname{#1}\apply@rg}   
\def\apply@rg#1,{\ifx\end#1\let\next=\relax
\else,\thecsname{#1}\let\next=\apply@rg\fi\next}

\def\citer@nge#1{\citedor@nge#1-\end-}  
\def\citer@ngeat#1\end-{#1}
\def\citedor@nge#1-#2-{\ifx\end#2\r@featspace#1 
  \else\citel@@p{#1}{#2}\citer@ngeat\fi}    
\def\citel@@p#1#2{\ifnum#1>#2{\errmessage{Reference range #1-#2\space is bad.}%
    \errhelp{If you cite a series of references by the notation M-N, then M and
    N must be integers, and N must be greater than or equal to M.}}\else%
 {\count0=#1\count1=#2\advance\count1 by1\relax\expandafter\r@fcite\the\count0,
  \loop\advance\count0 by1\relax
    \ifnum\count0<\count1,\expandafter\r@fcite\the\count0,%
  \repeat}\fi}

\def\r@featspace#1#2 {\r@fcite#1#2,}    
\def\r@fcite#1,{\ifuncit@d{#1}
    \newr@f{#1}%
    \expandafter\gdef\csname r@ftext\number\r@fcount\endcsname%
                     {\message{Reference #1 to be supplied.}%
                      \writer@f#1>>#1 to be supplied.\par}%
 \fi%
 \csname r@fnum#1\endcsname}
\def\ifuncit@d#1{\expandafter\ifx\csname r@fnum#1\endcsname\relax}%
\def\newr@f#1{\global\advance\r@fcount by1%
    \expandafter\xdef\csname r@fnum#1\endcsname{\number\r@fcount}}

\let\r@fis=\refis           
\def\refis#1#2#3\par{\ifuncit@d{#1}
   \newr@f{#1}%
   \w@rnwrite{Reference #1=\number\r@fcount\space is not cited up to now.}\fi%
  \expandafter\gdef\csname r@ftext\csname r@fnum#1\endcsname\endcsname%
  {\writer@f#1>>#2#3\par}}

\def\ignoreuncited{
   \def\refis##1##2##3\par{\ifuncit@d{##1}%
    \else\expandafter\gdef\csname r@ftext\csname r@fnum##1\endcsname\endcsname%
     {\writer@f##1>>##2##3\par}\fi}}

\def\r@ferr{\endreferences\errmessage{I was expecting to see
\noexpand\endreferences before now;  I have inserted it here.}}
\let\r@ferences=\references
\def\references{\r@ferences\def\endmode{\r@ferr\par\endgroup}}

\let\endr@ferences=\endreferences
\def\endreferences{\r@fcurr=0
  {\loop\ifnum\r@fcurr<\r@fcount
    \advance\r@fcurr by 1\relax\expandafter\r@fis\expandafter{\number\r@fcurr}%
    \csname r@ftext\number\r@fcurr\endcsname%
  \repeat}\gdef\r@ferr{}\endr@ferences}


\let\r@fend=\endpaper\gdef\endpaper{\ifr@ffile
\immediate\write16{Cross References written on []\jobname.REF.}\fi\r@fend}

\catcode`@=12

\citeall\refto      
\citeall\ref        %
\citeall\Ref        %

\def\D{{\cal D}}

\def\E{{\cal E}}
\def\p{\partial}
\def\la{\langle}
\def\ra{\rangle}
\def\ria{\rightarrow}
\def\x{{\bar x}}
\def\xh{{\hat x}}
\def\yh{{\hat y}}
\def\ph{{\hat p}}

\def\kh{{\hat k}}
\def\Qh{{\hat Q}}
\def\Kh{{\hat K}}
\def\Ph{{\hat P}}
\def\Xh{{\hat X}}

\def\s{{\sigma}}
\def\a{\alpha}

\def\U{\Upsilon}

\def\om{{\omega}}
\def\Tr{{\rm Tr}}
\def\ih{{ {i \over \hbar} }}

\def\au{{\underline \alpha}}

\centerline{\bf Approximate Decoherence of Histories}
\centerline{\bf and 't Hooft's Deterministic Quantum Theory}

\bigskip
\bigskip
\author J.J.Halliwell 
\vskip 0.2in
\affil
Theory Group
Blackett Laboratory
Imperial College
London SW7 2BZ
UK
\vskip 0.5in
\centerline {\rm Preprint Imperial/TP/00-1/5}
\vskip 0.1in
\centerline{\rm November, 2000}
\vskip 0.1in
\vskip 0.1in
\vskip 0.3in

\abstract{In the decoherent histories approach to quantum theory,
sets of histories are said to be decoherent when the decoherence
functional, measuring interference between pairs of histories, is
exactly diagonal. In realistic situations, however, only
approximate diagonality is ever achieved, raising the question of
what approximate decoherence actually means and how it is related
to exact decoherence. This paper explores the possibility that an
exactly decoherent set of histories may be constructed from an
approximate set by small distortions of the operators
characterizing the histories. In particular, for the case of
histories of positions and momenta, this is achieved by doubling
the set of operators and then finding, amongst this enlarged set,
new position and momentum operators which commute, so decohere
exactly, and which are ``close'' to the original operators. Two
derivations are given, one in terms of the decoherence functional,
the second in terms of Wigner functions. The enlarged, exactly
decoherent, theory has the same classical dynamics as the original
one, and coincides with the so-called deterministic quantum
theories of the type recently studied by 't Hooft. These results
suggest that the comparison of standard and deterministic quantum
theories may provide an alternative method of characterizing
emergent classicality. A side-product is the surprising result
that histories of momenta in the quantum Brownian motion model
(for the free particle in the high-temperature limit) are exactly
decoherent.}

\endtopmatter
\endpage

\head{\bf 1. Introduction}

How close to classical mechanics can quantum mechanics be? One of
the main aims of the decoherent histories approach is to
demonstrate the emergence of classical mechanics as an effective
theory, starting from the assumption that quantum mechanics is the
exact underlying theory [\cite{Gri,Omn,GeH1,GeH2,Har6}]. In such studies, the
effective classical theory almost always emerges in an approximate
way, rarely exact. The main reason for this is that decoherence,
the destruction of quantum interference, is almost always
approximate. What does approximate decoherence mean? What is the
nature of the histories that approximately decoherent histories
are an approximation to?

The aim of this paper is to explore the idea that approximate
decoherence of histories can be turned into exact decoherence by
suitable ``small'' modifications of the operators characterizing
the histories. In particular, histories characterized by fixed
values of coordinates and momenta $x,p$ are rendered exactly
decoherent by replacing $x,p$ with new coordinates and momenta
$X,P$ which {\it commute}.  This replacement, we show, is a valid
approximation provided that the original histories are
approximately decoherent. The new theory in terms of the commuting
variables $X,P$ has the same form as the so-called deterministic
quantum theories of the type recently studied by 't Hooft [\cite{Hoo}].

To set up the problem in more detail, we briefly review
the decoherent histories approach [\cite{Gri,Omn,GeH1,GeH2,Har2,Hal1}].
In the decoherent histories approach to quantum theory, probabilities
are assigned to histories of a closed system via the formula,
$$
p(\a_1, \a_2, \cdots \a_n) = {\rm Tr} \left( P_{\a_n}(t_n)\cdots
P_{\a_1}(t_1)
\rho P_{\a_1} (t_1) \cdots P_{\a_n} (t_n) \right)
\eqno(1.1)
$$
The projection operators $P_{\a}$ characterized the different
alternatives describing the histories at each moment of time.
The projectors satisfy
$$
\sum_{\a} P_{\a} =1, \quad \quad
P_{\a} P_{\beta} = \delta_{\a \beta} \ P_{\a}
\eqno(1.2)
$$
and the projectors appearing in (1.1) are in the Heisenberg picture,
$$
P_{\a_k}(t_k) = e^{i H(t_k-t_0)} P_{\a_k} e^{-i H(t_k-t_0)}
\eqno(1.3)
$$
Probabilities can be assigned to histories if and only if all histories
in the set obey the condition of {\it consistency}, which is that
$$
{\rm Re} D(\au , \au' ) = 0
\eqno(1.4)
$$
for $\au \ne \au' $. Here $\au$ denotes the string $\a_1, \cdots
\a_n$ and $D(\au , \au') $ is the decoherence functional,
$$
D(\au , \au') = {\rm Tr} \left( P_{\a_n}(t_n)\cdots
P_{\a_1}(t_1)
\rho P_{\a_1'} (t_1) \cdots P_{\a_n'} (t_n) \right)
\eqno(1.5)
$$
Loosely speaking, the decoherence functional measures
the amount of interference between pairs of histories.
It is observed in numerous examples involving
physical mechanisms for decoherence
that the imaginary
part of the decoherence functional often also vanishes
when the real part vanishes,
and it is therefore of interest to consider the stronger
condition of decoherence,
$$
D (\au , \au' ) = 0
\eqno(1.6)
$$
for $\au \ne \au'$. This condition may be shown to be related
to the existence of records -- projectors which may be
added to the very end of
the string of projectors which are perfectly correlated with
the earlier alternatives $\a_1, \cdots \a_n$, and are related
to the physical process of information storage
[\cite{GeH2,Hal5}].

In its application to physical interesting situations, therefore,
one of the first aims of the approach is to find out how the
decoherence condition (1.6) may come to be satisfied. This is often
accomplished, for example, by coupling the system of interest to an
environment and then tracing out the environment. Or more generally,
by some kind of coarse-graining procedure. However, as
indicated earlier, it is almost
universally observed in such situations that the condition (1.6) is
only satisfied approximately, not exactly. The degree to which this
condition is satisfied can be exceptionally good, by any standards
(see Refs.[\cite{DoH,McE}], for example),
but it is still nevertheless approximate.
Although to work with approximate decoherence seems very
reasonable physically, from a more rigorous point of view
it leaves a grey area in the formalism, since it is not
clear what the approximately decoherent histories
are an approximation to, if anything [\cite{DoK}]. It would be
highly desirable to find a more controlled way of moving
between approximate and exact decoherence.

As stated above, we shall show that there is a closely
related theory which is exactly decoherent and which,
under certain circumstances, approximately coincides
in its predictions with the approximately decoherent
theory.

We start with the observation that the generic lack of decoherence
of histories is due to the fact that operators at different times
generally do not commute. In the case of histories characterized by
projections onto positions, positions at different times can be
completely expressed in terms of $\ph$ and $\xh$ at the initial time,
so the non-decoherence is due to non-commutativity of the basic
canonical pair. Histories characterized by operators which do
commute at different times are exactly decoherent, as may be seen
from (1.6). (Histories of conserved quantities are important
examples of this type [\cite{HLM}]).

We now recall a very old result due to von Neumann, concerning
the non-commuting pair, $\ph$, $\xh$. Von Neumann showed that it
is possible to find a new pair of operators, $\ph'$, $\xh'$, say,
which do commute, and which are in some sense ``close'' to the
original pair [\cite{Neu}]. The key issue is then to explain what is meant by
``close''. This is obviously a rather subtle issue. Every
interesting quantum effect can be traced back to non-commuting
operators, so clearly there will be many situations in which this
replacement is a very poor approximation. The point, of
course, is that the measure of closeness depends on the context.
We are primarily interested situations which are almost classical
anyway, and in that case there is a chance that such an
approximation may be good.

This suggests the following approach to approximate decoherence.
We start with a decoherence functional which is approximately
diagonal. Replace the operators with commuting operators, thereby
achieving exact diagonality. The degree of closeness is then
measured by the amount that the probabilities for the histories
change on replacing the original operators with the commuting
operators. We expect this change to be small when the original set
of histories are approximately decoherent. Of course, {\it any}
set of histories can be made exactly decoherent in this way.
The point, however, is that we expect only histories which are
approximately decoherent in the first place will undergo a small
change in their probabilities through this procedure. Sets of
histories which are not, by any reasonable standard, close to
being decoherent, will suffer a large change in their
probabilities.

The von Neumann method above is one way of obtaining a commuting
set of operators, and there are probably many ways of achieving
similar results. Here, we will use a different method which is
perhaps easier and more physically insightful, but is also perhaps
more radical in that involves changing the fundamental theory one
is quantizing. Suppose we start with a non-commuting canonical
pair, $\ph, \xh$, for a single particle in one dimension, so
$$
[\xh, \ph ] = i \hbar
\eqno(1.7)
$$
Denote this system $A$, and now
adjoin to it an auxiliary
system, denoted $B$, identical to $A$, with canonical
pair, $\kh, \yh$, and consider the variables
$$
\hat X = \xh + \yh, \quad \hat Q = \half( \xh - \yh),
\quad \hat K = \half (\ph + \kh), \quad \hat P = \ph - \kh
\eqno(1.8)
$$
We now have the commutation relations
$$
[ \hat Q, \hat P ] = i \hbar, \quad [\hat X, \hat K ] = i \hbar
\eqno(1.9)
$$
All other commutators are zero, and in particular,
note that
$$
[\hat X,\hat P] = 0
\eqno(1.10)
$$
Classically, we could set $y = 0 = k $ identically, so
$ X = x $ and $ P = p $. Quantum mechanically, we cannot
do this, but we can see how close we can get.
Suppose we put system $B$ in a minimum uncertainty state
with $\la \yh \ra = 0 = \la \kh \ra $. Then
$$
\la \hat X \ra = \la \xh \ra, \quad \la \hat P \ra = \la \ph \ra
\eqno(1.11)
$$
but the higher moments of $\yh$ and $\kh$ are non-zero. This
indicates that the pair $\ph, \xh $ are equal to the commuting
pair $\hat P,\hat X $ up to ``quantum fluctuations''. More
precisely, a measure of the degree of closeness is indicated by
the relations
$$
\la (\Xh -\xh)^2 \ra \ \la ( \Ph - \ph)^2 \ra
= \la \yh^2 \ra \ \la \kh^2 \ra = { \hbar^2 \over 4}
\eqno(1.12)
$$
The issue is
then to determine to what extent and under what conditions these
fluctuations are significant. Clearly they will be significant
when quantum-mechanical effects are important, but it is
reasonable suppose that they won't be significant close to the
classical regime.

To use this scheme in the decoherent histories approach it
is useful to write down an action for the extended system,
so that we can use path integrals. Recall that what we
ultimately need to get decoherence of position histories
is that positions at different times need to commute.
We therefore require that the non-commuting
operators $\hat x_t$ and $\xh$ at different times are distorted
into commuting operators $\Xh_t$, $\Xh$,
which will guarantee exact decoherence. Since, in
any reasonable dynamics, $\Xh_t$ is a function of
$ \dot \Xh $ and $\Xh$, the relationship between
velocities and momenta (so far unspecified) must be such
that $ [\dot \Xh, \Xh ] = 0 $. With the standard
action, we would have $ K = m \dot X $, since $K$ is
defined to be the conjugate to $X$, but this clearly
will not work since $ [ \Kh, \Xh ] \ne 0 $. We must
instead arrange that $ P = m \dot X $. It is easily
seen that this is achieved using that action
$$
\eqalignno{
S &= \int dt \left[ \half m \dot x^2 - \half m \dot y^2 \right]
\cr
&= \int dt \ m \dot X \dot Q
&(1.13) \cr }
$$
(in the free particle case).
The classical solution for $X$ in the free particle case is
$$
X_t = X + t \dot X = X + {Pt \over m}
\eqno(1.14)
$$
On quantization, this implies that $[\Xh_t,\Xh]= 0$
as required.

The action for
the new variables $X$ and $Q$ has the
form of the action for the deterministic
quantum theory (hereafter denoted
DQT) discussed by 't Hooft [\cite{Hoo}]. For the more general case
of a particle in a potential, this action is,
$$
S = \int dt \left( m \dot Q \dot X - Q V'(X) \right) \eqno(1.15)
$$
This produces the classical equations of motion
$$
m \ddot X + V'(X) = 0
\eqno(1.16)
$$
so gives the same classical dynamics as the usual action. But the
quantum theory will generally be quite different, since there are
twice as many variables. Furthermore, there is a price to pay in
that the Hamiltonian for this theory is unbounded below, although
there is some chance that this problem may be rectified by fixing
the quantum state of the auxiliary system $B$. Nevertheless, this
theory does have properties to recommend it for the purposes
of this paper: it is exactly
decoherent, and its classical dynamics coincides with the dynamics
of the original theory.

The work of 't Hooft concerns the possibility that the
deterministic quantum theory is a new fundamental theory,
replacing the standard one [\cite{Hoo}]. The reproduction of
quantization-like effects (in particular, discrete spectra) is
argued to arise from dissipative effects in the underlying
classical theory [\cite{Hoo,Bla}], making use of the fact that the basic action (1.15)
is readily modified to include dissipation at a fundamental level,
$$
S = \int dt \left( m \dot Q \dot X -2 m \gamma Q \dot X - Q V'(X)
\right)
\eqno(1.16)
$$
The present work is not primarily concerned with promoting
this point of view, but rather, with finding what sort of
mathematical statements one can make about the relationship
between approximate and exact decoherence. The results do,
however, contribute to 't Hooft's programme, in that they show in
detail how the predictions of standard quantum theory and
deterministic quantum theory become indistinguishable as the
classical regime is approached.

The results of this paper are basically simple and in some
ways almost obvious: DQT reproduces classical predictions exactly,
and standard quantum theory reproduces classical predictions
approximately when approximate decoherence holds, hence it is no
surprise that the two theories approximately coincide. The main task of this
paper, however, is to show in detail exactly how this works out.

In Section 2 we discuss the quantization of systems described by the
action (1.15). We show that histories of $X$ are exactly decoherent
and that the predictions of the theory may be arranged to coincide
{\it exactly} with those of the classical theory.

In Section 3, we discuss the standard picture of approximate
decoherence of histories of a simple linear system, with decoherence
provided by coupling to a thermal environment.

The main result of this paper is contained in Section 4, where we
repeat the analysis of Section 3 but with the addition of an
identical auxiliary system with the wrong sign action. We verify
that histories of $X = x+y$ are exactly decoherent, as in Section 2,
but here complicated by the
presence of an environment. Most importantly, the environment
ensures that the exactly decoherent deterministic theory makes
predictions which are indeed very close to the predictions of the
standard theory with approximate decoherence.

In Section 5, we give an alternative account of the results of
Section 4, working with the Wigner function rather than the
decoherent histories approach. We show that the Wigner function of
the DQT is a good approximation to the Wigner function of the
standard quantum theory approach if there is an environment
present. The role of the environment in both Sections 4 and 5 is
seen to be, through its fluctuations, to smear out the positions and
momenta so that the distinction between $x,p$ and $X,P$ becomes
insignificant.

In Section 6, we consider a different issue related to
the general theme of exact decoherence. This is the observation that
there is, in fact, an exactly decoherent set of histories already
buried in the standard approach, in the much-studied quantun
Brownian motion model. Namely, histories of momenta in this model
are exactly decoherent (for the free particle with a high
temperature environment). This is a different sort of exact
decoherence, since it is related to total momentum conservation of
the system coupled to the environment, but it does not seem to have
been noticed previously.

In Section 7 we briefly consider the question of how the scheme may
extend to quantum systems not described by a simple canonical
pair obeying (1.7). We summarize and conclude in Section 8.

\head{\bf 2. Deterministic Quantum Theories}

We now consider the quantization of the DQT described
by the action (1.15).
The Hamiltonian is
$$
H = {1 \over m } P K + Q V'(X)
\eqno(2.1)
$$
where recall we have the fundamental commutation relations
$$
[\Qh,\Ph ] = i \hbar , \quad [\Xh, \Kh] = i \hbar
\eqno(2.2)
$$
Since $ [\Xh, \Ph ] = 0 $, we may quantize using a representation
in which the wave functions depend on $X$ and $P$,
$ \tilde \Psi = \tilde \Psi (X,P)$. (Note one could
instead work with the commuting pair $\Qh, \Kh $ and work
in a representation in which $\Psi = \Psi (Q,K)$).
We therefore make the replacements,
$$
\Qh = i \hbar { \partial \over \partial P}, \quad
\Kh = - i \hbar { \partial \over \partial X}
\eqno(2.3)
$$
Hence the Schr\"odinger equation is
$$
i \hbar { \partial \over \partial t }
\tilde \Psi (X,P,t) = \left( - { i \hbar \over m } P { \partial \over
\partial X} + i \hbar V'(X) { \partial \over \partial P} \right)
\tilde \Psi (X,P,t)
\eqno(2.4)
$$
The factors of $i$ and $\hbar$ drop out, giving the Sch\"odinger
equation a totally classical form,
$$
{ \partial \over \partial t }
\tilde \Psi (X,P,t) = \left( - { P \over m }  { \partial \over
\partial X} + V'(X) { \partial \over \partial P} \right)
\tilde \Psi (X,P,t)
\eqno(2.5)
$$
This is a classical
Liouville equation (although note that the wave function
is not necessarily real). The solution is
$$
\tilde \Psi (X,P,t) = \tilde \Psi (X_{-t}, P_{-t}, 0 )
\eqno(2.6)
$$
where $X_{-t}$, $P_{-t}$ are the (backwards evolved)
classical solutions with initial data $X$, $P$.

We now see why the quantum theory of this system may be
called deterministic.
First of all, since $[\Xh, \Ph ] = 0 $,
we may choose initial states which are
arbitrarily concentrated in both $P$ and $X$. Secondly,
there is no wavepacket spreading in the dynamics (2.6),
and the states therefore remain arbitrarily peaked
in $P$ and $X$. There is therefore no obstruction to
assigning definite values to $X$ and $P$
for all times. There is also no possibility
of interference because interference arises from
wavepacket spreading. Because of these properties,
the predictions of this quantum theory may be arranged
to {\it exactly} coincide with the classical theory.
Much of the above has already been noted by 't Hooft
[\cite{Hoo}].

In the decoherent histories approach, these features ensure
that the histories of fixed $X$ are exactly decoherent, not surprisingly.
We briefly sketch the proof of this using a path integral
representation of the decoherence functional. It is
$$
\eqalignno{
D(\a, \a') = \int_{\au} & \D X(t) \int_{\au'}\D X'(t)
\int \D Q(t) \D Q'(t) \ \exp \left( \ih S [ X,Q] - \ih S [X',Q'] \right)
\cr & \times
\ \Psi_0 (X_0, Q_0) \Psi^*_0 (X_0', Q_0')
&(2.7) \cr}
$$
The sum is over pairs of paths $X(t),Q(t)$ and $X'(t),Q'(t)$
where $X(t), X'(t)$ are constrained to pass through
a series of gates denoted by $\a,\a'$
(described in more detail in Section 3), and $Q(t),Q'(t)$
are unrestricted.
The paths meet at the final point $t=t_f$, hence
$$
X_f = X_f', \quad Q_f =Q_f'
\eqno(2.8)
$$
After an integration by parts, the action (1.15) may be
written,
$$
S[X,Q] = - \int dt \ Q \left( m \ddot X + V'(X) \right)
+ m Q_f \dot X_f - m Q_0 \dot X_0
\eqno(2.9)
$$
and similarly,
$$
S[X',Q'] = - \int dt \ Q' \left( m \ddot X' + V'(X') \right)
+ m Q_f \dot X_f' - m Q_0' \dot X_0'
\eqno(2.10)
$$
where the final conditions (2.8) have been used.
Now consider the functional integral over $Q$. In a time-slicing
definition of this path integral, we may split the functional
integral into an integral of the initial values $Q_0$, $Q_0'$,
the final value $Q_f = Q_f'$, and the values on the interior
slices. The $Q(t)$
and $Q'(t)$ in the
integrands in (2.9), (2.10), sit on the interior slices only,
and
integrating them out pulls down delta-functions on the
equations of motion. Furthermore, the integral over
$Q_f = Q_f'$ pulls down a delta-function $\delta (\dot X_f
-\dot X_f')$. Hence, we obtain,
$$
\eqalignno{
D(\a, \a') = & \int_{\au} \D X(t) \int_{\au'} \D X'(t)
\ \int dQ_0 dQ_0'
\cr
& \times \ \delta \left[ m \ddot X - V'(X) \right]
\ \delta \left[ m \ddot X' - V'(X') \right]
\cr
& \times \ \delta (\dot X_f' - \dot X_f)
\ \exp \left( { i m \over \hbar} ( Q_0' \dot X_0' - Q_0 \dot X_0)
\right)
\cr
&
\times \ \Psi_0 (X_0, Q_0) \Psi^*_0 (X_0', Q_0')
&(2.11) \cr}
$$
Because of the delta-functions on the equations of motion
the sums over paths $X(t)$ and $X'(t)$ take contributions
only from histories satisfying the classical equations
of motion. But we also have the final condition
$X_f = X_f'$, together with the delta-function
in (2.11) which ensures that $\dot X_f = \dot X_f'$.
Therefore $X(t)$ and $X'(t)$ satisfy the same
second order equation and the same final conditions.
It follows that $X(t) = X'(t)$ in this path integral
and therefore there is exact decoherence.

The integral over $Q_0$ and $Q_0'$ performs a
Fourier transforms of initial wave function to the
representation $\tilde \Psi (X,P) $ used earlier,
$$
\tilde \Psi (X,P) = \int dQ \ e^{ - \ih PQ} \ \Psi (X,Q)
\eqno(2.12)
$$
and we find that the probabilities for the histories
are given by
$$
p(\a) = \int_{\a} \D X(t) \ \delta
\left[ m \ddot X + V'(X) \right]
\left| \tilde \Psi (X_0, m \dot X_0 ) \right|^2
\eqno(2.13)
$$
This is precisely the expected result for
a classical deterministic theory with
probability for initial conditions given by
$ | \tilde \Psi (X,P) |^2 $.

Finally, it is of interest to compare the initial phase
space distribution $ | \tilde \Psi (X,P) |^2 $ with
the Wigner function, which often crops up in this
sort of decoherence functional calculation [\cite{GeH2,Hal6}]. The
Wigner function is defined in terms of the wave function
$ \Psi (X,Q) $ by [\cite{Wig}]
$$
\eqalignno{
W(K,X,P,Q) = {1 \over (2 \pi \hbar)^2 }
& \int d \xi_1 d \xi_2
\ e^{ - \ih K \xi_1 - \ih P \xi_2 }
\cr \times &
\ \Psi (X+ \half \xi_1, Q + \half \xi_2 )
\ \Psi^* ( X - \half \xi_1, Q - \half \xi_2 )
&(2.14) \cr }
$$
Inserting the expression for $\Psi (X,Q)$ in
terms of its Fourier transform $ \tilde \Psi (X,P)$
(the inverse of (2.12)), it is easily shown that
the reduced Wigner function $ \tilde W (X,P) $
is
$$
\tilde W( X,P) = \int dK d Q \ W (K,X,P,Q) =
| \tilde \Psi (X,P) |^2
\eqno(2.15)
$$
which is the intuitively expected result.

\head{\bf 3. Approximate Decoherence in the Standard Picture}

We now briefly review the approximate decoherence of
position histories in standard quantum theory (denoted SQT).
We consider a single particle in a potential $V(x)$
linearly coupled to a large environment of
harmonic oscillators in an initial thermal state
with temperature $T_A$.
The action for this system is
$$
S [x,q_n]  = \int dt \left[ \half m \dot x^2 - V(x)\right]
+ \sum_n \int dt \left[ \half m_n \dot q_n^2 - \half m_n \om_n^2
q_n^2 - c_n q_n x \right]
\eqno(3.1)
$$
and the Hamiltonian is
$$
H = { p^2 \over 2m } + V(x)
+ \sum_n \left[ { p_n^2 \over 2 m } + \half m_n \om_n^2 q_n^2
+ c_n q_n x \right]
\eqno(3.2)
$$
This model, the quantum Brownian motion model,
has been considered many times elsewhere [\cite{FeV,CaL,QBM}],
especially in the context of decoherence [\cite{Zur1}]
(see also the older related work Ref.[\cite{JoZ}]).
We will describe it only in outline, quoting
required results where necessary.

After tracing out the environment variables,
the decoherence functional is
$$
\eqalignno{ D(\au, \au') = & \int {\cal D} x(t) {\cal D} x'(t)
\prod_{k=1}^n \U ( x(t_k) - \x_k ) \U ( x'(t_k) - \x_k' ) \cr &
\times \ \exp \left( \ih \int_0^{\tau} dt \left[ \half m \dot x^2
- V(x) - \half m
\dot x'^2 + V(x')
\right] \right)
\cr & \times \ F[ x(t), x'(t)] \ \rho_A (x_0, x_0')
&(3.3)\cr}
$$
Here, we use $\au$ to denote the string $\x_1, \x_2, \cdots \x_n$.
The window functions $\U$ restrict the paths to pass through
gates of width $\Delta $ centred about points $\x_1, \x_2, \cdots$
at times $t_1, t_2 \cdots t_n$ in a total time interval $[0,\tau]$.
The only leftover of the environment is the
influence functional,
$$
F[x(t),x'(t)] = \exp \left( \ih W[x(t), x'(t)] \right)
\eqno(3.4)
$$
where $ W[x(t),x'(t)] $ is the Feynman-Vernon influence functional phase,
$$
\eqalignno{
W[x(t),x'(t)] = & -
\int_0^\tau dt \int_0^t ds \ [ x(t) - x'(t) ] \ \eta (t-s) \ [ x(s) + x'(s) ]
\cr &
+ i \int_0^\tau dt \int_0^t ds \ [ x(t) - x'(t) ] \ \nu(t-s) \ [ x(s) - x'(s) ]
&(3.5) \cr }
$$
Full details of the kernels $\eta$ and $\nu$ may be found elsewhere
[\cite{FeV,CaL,AnHa,HPZ}].
They are in general non-local in time, but simplify enourmously
in the Fokker-Planck limit (high temperature and
a continuum of oscillators with a high frequency cut-off)
in which,
$$
\eqalignno{
W = & - \int_0^\tau dt \ m \gamma (x-x') (\dot x + \dot x')
- \int_0^\tau  dt \ \delta \omega^2 \ (x^2 -{x'}^2)
\cr &
+ {2 M \gamma k T_A \over \hbar } i \int_0^\tau dt
\ (x-x')^2
&(3.6) \cr }
$$
In what follows, to make the exposition clearer,
we will work entirely in this limit. (It is readily
verified that the following calculations can be
carried out with the fully general form (3.5), but
the expressions are much more cumbersome).

From (3.6) one can see that
the real part of $W[x(t),x'(t)]$ contributes a dissipative part to the
effective equations of motion, and also a renormalization
$\delta \omega^2$ to the frequency. We shall assume that the latter
has been absorbed into the potential $V(x)$.
The imaginary part produces
the decoherence, since it suppresses differing values of
$x$ and $x'$. Since the projectors coarse-grain the paths into regions
of size $\Delta$, distinct histories have $| x-x' | $ greater
than $\Delta$. The condition for approximate
decoherence is therefore loosely given by
$$
2 m \gamma k T_A \tau \Delta^2 >> \hbar^2
\eqno(3.7)
$$
hence is satisfied for sufficiently large
temperature. The imaginary part of $W[x(t),x'(t)]$
also produces fluctuations about the
effective classical equations of motion.

Given approximate decoherence,
we may take the probabilities for histories
to be given, to a good approximation,
by the diagonal elements of the decoherence functional.
The resulting expression is most easily evaluated using
the sum and difference coordinates,
$$
\xi = x - x', \quad u = \half (x + x')
\eqno(3.8)
$$
and we obtain for the probabilities for histories,
$$
\eqalignno{ p(\au)  = & \int {\cal D} u(t) {\cal D} \xi (t)
\prod_{k=1}^n \U ( u(t_k) + \half \xi (t_k)  - \x_k )
\U ( u(t_k) - \half \xi (t_k)  - \x_k ) \cr &
\times \ \exp \left( \ih \int dt \left[ m \dot u \dot \xi
- 2 m \gamma \dot u \xi - V ( u + \half \xi ) + V ( u - \half \xi)
\right] \right)
\cr
& \times \ \exp \left( - {2 m \gamma k T_A \over \hbar^2}
\int dt \ \xi^2 \right) \ \rho_A (u_0 + \half \xi_0, u_0 - \half \xi_0)
&(3.9)\cr}
$$
Consider the functional integral over $\xi$. It is Gaussian
except for the appearance of $\xi$ in the window functions
$\U$ and in the potential $V$.
However, the contribution from $\xi$ is very tightly
concentrated around $\xi = 0$. We there expect to be able
to drop the $\xi $ terms in the window functions, in comparison
to $u$, and also to use a small $\xi$ approximation in
the potential,
$$
V ( u + \half \xi ) - V ( u - \half \xi )
= \xi V'(u) + {1 \over 24} \xi^3 V^{\prime \prime \prime} (u)
+ \cdots
\eqno(3.10)
$$
Dropping the order $\xi^3$ term (shown here only for comparison
with later results),
the integral in the imaginary part of the exponential
may be integrated by parts yielding
$$
- \int dt \ \xi \ \left[ m \ddot u  + 2 m \gamma \dot u
+ V'(u)
\right] - \dot u_0 \xi_0
\eqno(3.11)
$$
where we have used the fact that $x=x'$ at the final time so
$\xi_f = 0$. In a skeletonized version of the path integral,
the integrand in (3.11) does not involve $\xi_0$, only the values
of $\xi$ on the internal time slices. The integral over $\xi_0$
with the boundary term from (3.11)
therefore effectively performs the Wigner transform of the initial
density matrix,
$$
W (p, u_0 ) = { 1 \over 2 \pi \hbar}
\int d \xi_0 \ e^{ - \ih p \xi_0 }
\ \rho_A (u_0 + \half \xi_0, u_0 - \half \xi_0)
\eqno(3.12)
$$
And carrying out the $\xi$ integral on the internal
times slices as well,
we therefore obtain,
$$
\eqalignno{ p(\au)  = & \int {\cal D} u(t)
\prod_{k=1}^n \U ( u(t_k) - \x_k )
\cr &
\times \exp \left( - { 1 \over 8 m\gamma k T_A }
\int dt \left[ m \ddot u + 2m \gamma \dot u + V'(u) \right]^2
\right)
\ W ( m \dot u_0, u_0 )
&(3.13) \cr}
$$
This is the desired result. A simple expression for the
probability for histories of positions. It is peaked about
classical evolution with dissipation, with thermal fluctuations
about that motion, and with the initial data weighted by
the Wigner function of the initial state. (The Wigner function
is not always positive, but a closer analysis of this sort of
expression [\cite{Hal6}] reveals that the Wigner function is
effectively smeared in such a way that it is positive).

Eq.(3.13) was derived
under essentially one approximation: that the contribution
from paths with large values of $\xi = x-x'$ could
be neglected. This meant firstly, that the approximate
decoherence could be taken as essentially exact.
Secondly, that we could drop the $\xi$ terms in the window
functions in (3.9) and the higher powers of $\xi$
in the expansion of the potential (3.10), so that
we could carry out the $\xi$ integration.

\head{\bf 4. Comparison with the Exactly Decoherent Deterministic
Quantum Theory}

The formula (3.13) bears a close resemblance to Eq.(2.13), the
probabilities for histories in the exactly decoherent DQT.
There are, however, three differences. First, (3.13) has dissipation
in the equations of motion but (2.13) does not, but this is
easily fixed by the trivial generalization of (2.13) to
the case of the dissipative action (1.17). Second, (2.13)
has a delta-function peak about the equation of motion, whilst
(3.13) has only a Gaussian peak, due to the thermal fluctuations.
This Gaussian peak becomes sharper as the mass of the particle
increases. Moreover, the difference between the two types
of peak will not be noticed if the width of the projections in
(3.13) are much greater than the width of the Gaussian,
Third, (3.13) has a (not necessarily positive) Wigner
function weighting its initial conditions, whilst (2.13) has
a positive weight function. But given that the fluctuations
tend to smear $W$ so as to be positive anyway (as will be
discussed at greater length below), for a wide variety of initial
states it ought to be possible to choose an initial state in (2.13)
to give essentially the same results as (3.13).

Of the above differences, the most important one is the
delta-function versus Gaussian peak. We therefore conclude that as
long as the particle is sufficiently massive to substantially
resist the effects of thermal fluctuations, the exactly decoherent
DQT of Section 2 approximately reproduces the probabilities of the
approximately decoherent histories of standard quantum theory
described above. This is our first result on the closeness of DQT
and standard quantum theory.

The above result applies, however, only to the case when the mass
of the particle is sufficiently large to resist thermal
fluctuations. It does not apply to the case where there is
approximate decoherence but the fluctuations about classical
deterministic behaviour are not small, as in the case of small
mass. The most general effective theories emerging from an
underlying quantum theory are classical stochastic theories,
perhaps with large fluctuations. We therefore need to generalize
our comparison of DQT and standard quantum theory to this case,
and this turns out to be somewhat more complicated. It requires
comparing the quantum Brownian motion model of Section 3 to a DQT
including an environment to provide fluctuations.

We have seen for a simple linear system with action $S[x]$, a
closely related DQT may be constructed using the action $ S = S[x]
- S[y] $ and by focusing on the variable $ X = x+y$. The coupling
to an environment, as in Eq.(3.1), requires a reconsideration of
the question of how to construct the related DQT. On the basis of
what we have seen so far -- that the DQT is obtained by doubling
what we already have -- it seems natural to double up both the
system and the environment. Whilst this in fact turns out to be
correct, one might wonder whether it would be possible to obtain
exact decoherence by the simpler procedure of doubling the system
alone. As we shall see, however, the dissipative terms induced by
the environmental interactions prevent this from working properly.
We therefore do indeed need to double both system and environment.

One can imagine a number of different ways of proceeding at
this point. For example, one could extend the analysis of Section
2 to include coupling to a thermal environment and then
repeat the steps leading to Eq.(2.13). This would, however,
involve getting into unnecessary detail about the
environment dynamics and initial state. We will instead
stay as close as possible to the calculation of Section 3,
in which all the environment dynamics are concisely summarized
in the influence functional.

Consider therefore the same calculation as in Section 3 but
with both system and environment doubled up.
For simplicity, we first concentrate on the case of
a linear system with $V(x) = \half m \om^2 x^2 $.
We therefore consider system $A$ with coordinates $x$
coupled to its environment with temperature $T_A$,
as before, with the auxiliary system $B$ and its
environment, with temperature $T_B$ (which, we shall see,
does not have to be the same as $T_A$).
Following the general scheme, we consider histories
specified by fixed values of $X = x+y$. After tracing
out both environments, the decoherence functional is
$$
\eqalignno{ D(\au, \au') = & \int {\cal D} x(t) {\cal D} x'(t)
\D y(t) \D y'(t)
\prod_{k=1}^n \U ( x(t_k) + y(t_k)- \x_k )
\U ( x'(t_k) + y'(t_k) - \x_k' )
\cr & \times
\ \exp \left( \ih \int dt \left[ \half m \dot x^2
-\half m \omega^2 x^2 - \half m
\dot x'^2 + \half m \omega^2 x'^2 \right] \right)
\cr & \times
\ \exp \left( \ih \int dt \left[ -\half m \dot y^2
+\half m \omega^2 y^2 + \half m
\dot y'^2 - \half m \omega^2 y'^2 \right] \right)
\cr & \times
\ F_A[ x(t), x'(t)]
\ F^*_B[ y(t), y'(t)]
\ \rho_A (x_0, x_0')
\ \rho_B (y_0, y_0')
&(4.1)\cr}
$$
(where note that the effect of the wrong sign action in the
auxiliary system $B$ effectively gives the complex conjugate of
the influence functional). We will confirm that this is exactly
decoherent and compute the probabilities for histories. Note that
Eq.(4.1) and the corresponding approximately
decoherent expression (3.3) are almost identical: if
the projections in (4.1) were onto values of $x,x'$, rather than
$X=x+y$, $X'=x'+y'$, then all the $y$, $y'$ terms could be
entirely integrated out yielding (3.3).

The path integral is most easily evaluated by changing
variables from $(x,y)$ to $ (X,y)$ (and similarly for
the primed variables). In these coordinates, and writing
out the influence functional explicitly,
it reads,
$$
\eqalignno{ D(\au, \au') = & \int {\cal D} X(t) {\cal D} X'(t)
\D y(t) \D y'(t)
\prod_{k=1}^n \U ( X(t_k)- \x_k )
\U ( X'(t_k) - \x_k' )
\cr & \times
\ \exp \left( \ih \int dt \left[ \half m \dot X^2
-\half m \omega^2 X^2 - \half m
\dot X'^2 + \half m \omega^2 X'^2 \right] \right)
\cr & \times
\ \exp \left( \ih \int dt \left[ - m \dot y \dot X
+  m \omega^2 y X + m \dot y' \dot X'
- m \omega^2 y' X' \right] \right)
\cr & \times
\ \exp \left( \ih \int dt \left[ - m \gamma ( X - X') ( \dot
X + \dot X' )
\right. \right.
\cr & \quad \quad \quad \quad \quad \quad \quad \quad
\left. \left.
+ m \gamma ( y-y') (\dot X + \dot X')
+ m \gamma ( \dot y + \dot y') (X - X')
\right] \right)
\cr & \times
\ \exp \left( - { 2 m \gamma k T_A \over \hbar^2}
\int dt (X -X' - y +y')^2 - { 2 m \gamma k T_B \over \hbar^2}
\int dt (y-y')^2 \right)
\cr & \times
\ \rho_A (X_0 - y_0, X_0'-y_0')
\ \rho_B (y_0, y_0')
&(4.2) \cr}
$$
Recall that in Section 2, exact decoherence in the deterministic
model was obtained as a result of the action
being linear in one of the variables, hence yielding a
delta-function on integration. In this case, note that the exponent
is linear in the variable $y+y'$. So introduce new coordinates
$$
Y = \half ( y+y'), \quad v = y-y'
\eqno(4.3)
$$
and note that
$$
y X - y' X' = Y (X-X') + \half v \ (X+X')
\eqno(4.4)
$$
The $y$ terms in the second and third exponential in (4.2)
therefore become
$$
\eqalignno{
\int dt & \left[ - m \dot y \dot X
+  m \omega^2 y X + m \dot y' \dot X'
- m \omega^2 y' X'
\right.
\cr
& \quad \quad \quad \quad \left.
+ m \gamma ( y-y') (\dot X + \dot X')
+ m \gamma ( \dot y + \dot y') (X - X')
 \right]
\cr &
= \int dt \left[ - m \dot Y ( \dot X - \dot X') - \half m \dot v
\ (\dot X + \dot X') + m \om^2 Y (X-X') + \half m \om^2 v
\ (X + X')
\right.
\cr & \quad \quad \quad \quad \left.
+ m \gamma v \ ( \dot X + \dot X') + 2 m \gamma \dot Y
( X - X')
\right]
&(4.5) \cr}
$$
As advertized, the exponential in the path integral is
now entirely linear in $Y$, and, after an integration by
parts in (4.5),
$Y$ may be integrated
out on the interior slices to produce a delta function
on configurations satisfying the equation
$$
\ddot X - \ddot X' - 2 \gamma (\dot X - \dot X')
+ \om^2 (X - X') =0
\eqno(4.6)
$$
This is the anti-damped dissipative equation for $X-X'$,
but this
does not matter since it is not the effective equation
of motion (derived below). The integration by parts in
(4.5) also produces the boundary terms,
$$
- \left[ m Y (\dot X - \dot X') + 2m \gamma Y (X - X')
\right]_0^{\tau}
\eqno(4.7)
$$
As in Section 2, the integration over $Y_f$ produces
a delta function enforcing $\dot X_f = \dot X_f'$,
and since we also have $ X_f = X_f'$, the
solution to (4.6) is therefore $X(t) = X'(t)$ identically, for all $t$.
We therefore have exact decoherence, as expected. It follows
that the other boundary terms in (4.7) vanish (since they
are proportional to $X-X'$).

We may now compute the probabilities for histories. With
$X(t) = X'(t)$ throughout, we now have
$$
\eqalignno{ p (\au) = & \int {\cal D} X(t) \ \D v (t) \ dy_0  dy_0'
\prod_{k=1}^n \U ( X(t_k)- \x_k )
\cr & \times
\exp \left( \ih
\int dt \left[ - m \dot v \dot X
+ 2 m \gamma v \dot X  +  m \om^2 v X  \right] \right)
\cr & \times
\ \exp \left( - { 2 m \gamma k (T_A +T_B) \over \hbar^2}
\int dt \ v^2 \right)
\cr & \times
\ \rho_A (X_0 - y_0, X_0-y_0')
\ \rho_B (y_0, y_0')
&(4.8) \cr}
$$
The $v$ integral may now be carried out, and, noting that
there is also a boundary term coming from the integration
by parts of the $m \dot v \dot X $ term, we get
$$
\eqalignno{ p (\au) = & \int {\cal D} X(t)
\prod_{k=1}^n \U ( X(t_k)- \x_k )
\cr & \times
\exp \left( - { m \over 8 \gamma k T'}
\int dt \left[ \ddot X + 2 \gamma \dot X + \om^2 X
\right]^2 \right)
\cr & \times
\ \int dy_0 dy_0' \ e^{ \ih m (y_0 - y_0') \dot X_0 }
\ \rho_A (X_0 - y_0, X_0-y_0')
\ \rho_B (y_0, y_0')
&(4.9) \cr}
$$
where $T' = T_A + T_B$. Now consider the last part of this
expression, the $y_0, y_0'$ integral
involving the initial state. We now choose
$\rho_B$ to be the ground state of the
harmonic oscillator. If we also let $y_0 \ria - y_0$,
$ y_0' \ria - y_0'$, followed by the transformation,
$y_0 \ria y_0 - X_0 $
and $y_0' \ria y_0' - X_0 $, then this integral becomes
$$
\ \int dy_0 dy_0' \exp \left( - \ih m (y_0 - y_0') \dot X_0
- {(y_0 -X_0)^2 \over 4 \s^2} - { (y_0' - X_0)^2 \over 4 \s^2}
\right)
\ \rho_A ( y_0, y_0')
\eqno(4.10)
$$
This is clearly just the average of the initial state $\rho_A$
in a coherent state $ |p,q\ra $ with $p= m \dot X_0$, $q=X_0$.
Hence the final expression is
$$
\eqalignno{ p (\au) = & \int {\cal D} X(t)
\prod_{k=1}^n \U ( X(t_k)- \x_k )
\cr & \times
\exp \left( - { m \over 8 \gamma k T'}
\int dt \left[ \ddot X + 2 \gamma \dot X + \om^2 X
\right]^2 \right)
\cr & \times
\la m \dot X_0, X_0 | \rho_A | m \dot X_0, X_0 \ra
&(4.11) \cr}
$$
This is the desired result: the probability for histories
of $X$ for the exactly decoherent deterministic theory
with an environment.

The main issue now is to compare this result with
Eq.(3.13) derived using standard quantum theory under
the conditions of approximate decoherence. Eq.(4.11) is clearly
a much better approximation to (3.13) than (2.13) was.
Eq.(4.11) has the desired dissipation term (although here
it comes from the environment, and not from the action
(1.17)). Most importantly it has thermal fluctuations.
The temperature in (4.11) is $T' = T_A +T_B$, versus
a temperature $T_A$ in (3.13), but this difference is
clearly negligible if we choose $T_B << T_A$.

The only significant difference between (4.11) and (3.13)
is the appearance of the explicitly positive
weight on initial data, $ \la p,q | \rho_A | p,q \ra $, in (4.11),
versus the Wigner function $W(p,q)$ in
(3.13). The two objects are, however, close. $ \la p,q | \rho_A |
p,q \ra $ is readily shown to be equal to the  Wigner function of
$\rho_A$ but smeared over an $\hbar$-sized region of phase space.
Moreover, the subsequent evolution of the system renders the
difference between these two objects negligible, since the thermal
fluctuations produce a smearing in phase space which becomes much
greater than $\hbar$ on a very short time scale [\cite{HaZ,Hu}].
The probabilities of the DQT and the approximately
decoherent standard quantum theory are therefore very close.

The physical picture is as follows. We have proposed switching
from non-commuting operators $\xh,\ph$ to commuting ones $\Xh,\Ph$
differing from the original ones by ``quantum fluctuations''. The
key point is that in the presence of the environment, the system
also suffers thermal fluctuations which are typically much larger
than the quantum fluctuations in $\Xh-\xh$ and $\Ph-\ph$. The
difference between the two sets of operators is therefore
negligible, and we may reasonably consider the two theories as
``close''.

We now consider some finer points of this derivation.
Consider first the issue of why we need to include the environment
of the auxiliary system $B$. As stated, this has to do with the
dissipative term. The question is what would happen if we drop the
environment of $B$. It is easy to see that dropping the
fluctuation term for $B$'s environment does no harm. In fact it
improves things, since it is the same as setting $T_B = 0$ so we
no longer need the condition $T_A >> T_B$. On the other hand,
dropping the dissipative terms for $B$ is equivalent to including a
term proportional to
$$
m \gamma ( y - y') (\dot y + \dot y') = 2 m \gamma v \dot Y
\eqno(4.12)
$$
in the exponent in (4.2). On carrying out the integral
over $Y$, this produces a term proportional to $v$ on the
right-hand side of (4.6). The key point now is that
the solution to this equation is no longer $X(t) = X'(t)$
identically. Therefore exact decoherence is destroyed.
Hence the presence of the dissipative term is required.

A possible difficulty of having to include a second environment is
that its effects may become significant at low temperatures. We have
concentrated here on the high temperature regime, but in standard
quantum theory there is some decoherence at low temperatures,
including zero temperature (although this does not seem to have been
very extensively studied in the literature [\cite{Hal5,LowT}]). (At
low temperatures  note also that the fully non-local form of the
influence functional (3.5) must be used.) In this regime it becomes
less obvious that the DQT is close to the predictions of SQT.

At all temperatures, standard quantum theory, after approximate
decoherence, is approximately equivalent to a classical but
stochastic theory (described by (3.13) for example), consisting of
deterministic evolution according to classical equations of motion
with dissipation, with thermal fluctuations about that motion.
This description is still good even if the fluctuations are not
small. The DQT also leads to a description in terms of
fluctuations about deterministic evolution, but the presence of
{\it two} environments means that the fluctuations are not the
same in general as the fluctuations in the SQT case -- they are
larger, as evidence by the presence of the temperature $T_A + T_B$
in (4.11). They are approximately the same if $T_A >> T_B$, but
they will be different if both $T_A$ and $T_B$ are the same order
of magnitude. Hence, SQT and the DQT are generally not
approximately the same in their predictions for low temperature
environments, since the fluctuations in the DQT case are
significantly larger.

At least, that is the conclusion on the basis of the approach of
this section, involving doubled environments. It does not rule out
the possibility that another type of DQT might approximately
reproduce the predictions of standard quantum theory at low
temperatures. Indeed, if the mass of the particle is very large,
Eq.(2.13) with a dissipative term will do the job moderately well
(as discussed at the beginning of this section). Still, the
analysis of this paper leaves space for a more thorough discussion
of the connection between DQT and SQT in the low temperature
regime.

Finally, consider the case of non-linear systems.
When a more general potential
is present, we need to replace the potential
terms in $S[x] - S[y]$ with $ (x-y) V'(x+y) $
(to coincide with the action (1.15)).
This means of course that the systems $A$ and $B$
are now coupled whereas previously they were not.
It is readily shown that the analysis goes through
in a very similar way with a term $V'(X)$ in
the final result (4.11) in place of $m \om^2 X $.
General potentials are fully treated in the alternative
formulation in the next section.

\head{\bf 5. A Wigner Function Formulation.}

We have examined the relationship between SQT and DQT by
comparing the probabilities for histories of the two theories,
when SQT is approximately decoherent. This still leaves, however,
a certain amount of vagueness in a statement about the
relationship between approximate and exact decoherence, since the
probabilities from SQT are still only approximately defined due to
imperfect decoherence. A perhaps more precise way of comparing the
predictions of standard quantum theory with the deterministic one
is to compare the density operator of standard quantum theory with
the reduced density operator of DQT after the extra variables
($K,Q$, etc.) have been traced out. This we now do. We will in
fact work with the Wigner function [\cite{Wig}], rather than the
density operator, but this is essentially the same since they are
related by a simple Fourier transform.

The system $A$ plus its environment has Hamiltonian (3.2)
and is described by a Wigner function $ W(p,x,p_n,q_n)$ obeying
the equation
$$
{\p W \over \p t} =\{ H, W \} + D W
\eqno(5.1)
$$
where $\{\ , \ \} $ is the usual Poisson bracket and
$ D $ is an operator acting on phase space,
$$
D = \sum_{n=1}^{\infty} { (-1)^n \over 2^{2n} } {1 \over (2n+1)!}
{d^{2n+1} V (x) \over dx^{2n+1} }
{ \p^{2n+1}  \over \p p^{2n+1} }
\eqno(5.2)
$$
Explicitly,
$$
\eqalignno{
{ \p W \over \p t }
=&- { p \over m } {\p W \over \p x} + V'(x) { \p W \over \p p}
+ D W
\cr
& + \sum_n \left[ - { p_n \over m_n } { \p W \over \p q_n}
+ m_n \om_n^2 q_n { \p W \over \p p_n}
+ c_n x { \p W \over \p p_n}
+ c_n q_n { \p W \over \p p}
\right]
&(5.3) \cr }
$$
This equation describes the exact dynamics of the system $A$
coupled to its environment.
Assuming a factored initial state between
system and environment, and with a thermal initial state
with temperature $T_A$ for the environment,
the environment coordinates may be traced out, and
an equation for the reduced Wigner
function for the system only $ \bar W(x,p)$
may be derived. This is in general a non-Markovian
equation, whose explicit form is only readily obtained
for linear systems [\cite{HPZ,HaT}].
But in the Fokker-Planck limit
(used in the previous Section) it has
the form
$$
{\p \bar W \over \p t }
= - { p \over m } {\p \bar W \over \p x} + V'(x) { \p \bar W \over \p p}
+ 2 { \p \over \p p } ( p \bar W ) + 2 m \gamma k T_A { \p^2 \bar W \over \p p^2}
+ D \bar W
\eqno(5.4)
$$
It is well-known that the diffusion term spreads out the Wigner function
so that the higher derivative terms $D \bar W$ may be neglected
[\cite{HSZ,PaZ}].
Furthermore, the Wigner function also becomes positive after a very short
time [\cite{HaZ}]. It may therefore be regarded, approximately, as a classical phase
space distribution function. This is the usual account of the approximate
emergence of classical behaviour using the Wigner function
or density operator, paralleling the discussion of Section 3.

We now compare this to the Wigner function description of the
deterministic quantum theory, which we know to be exactly
decoherent,
paralleling the derivation of Section 4.
The action for the deterministic theory coupled to an environment is
$$
\eqalignno{
S = \int dt & \left[ m \dot Q \dot X - Q V'(X) \right]
\cr & + \sum_n \int dt \left[ m_n \dot Q_n \dot X_n - m_n \om_n^2 Q_n X_n
- c_n Q_n X - c_n X_n Q \right]
&(5.5) \cr}
$$
where the coordinates are related to the coordinates
$x,y$ etc. by
$$
X = x+y, \quad Q = \half (x-y), \quad X_n = q_n + \tilde q_n,
\quad Q_n = \half (q_n - \tilde q_n )
\eqno(5.6)
$$
In the linear case, the action (5.5) is of the form
$$
S = S [x,q_n] - S[y, \tilde q_n]
\eqno(5.7)
$$
The Hamiltonian is
$$
H = { 1 \over m } P K + Q V'(X)
+ \sum_n \left[ {1 \over m_n } P_n K_n + m_n \om_n^2 Q_n X_n
+ c_n Q_n X + c_n X_n Q \right]
\eqno(5.8)
$$
where $P_n, K_n$ are the momenta conjugate to $Q_n, X_n$
respectively. The Wigner function for this system
$ W= W(K,X,P,Q,K_n,X_n,P_n,Q_n) $ obeys the evolution
equation
$$
\eqalignno{
{ \p W \over \p t} =&  - { K \over m } { \p W \over \p Q}
- { P \over m} { \p W \over \p X} + V'(X) { \p W \over \p P}
+ Q V^{\prime\prime} (X) { \p W \over \p K}
+ \tilde D W
\cr &
+ \sum_n \left[
- {K_n \over m_n} {\p W \over \p Q_n}
- {P_n \over m_n} { \p W \over \p X_n}
+ m_n \om_n^2 X_n { \p W \over \p P_n}
+ m_n \om_n^2 Q_n { \p W \over \p K_n} \right]
\cr &
+ \sum_n c_n \left[ X { \p W \over \p P_n}
+ Q { \p W \over \p K_n} + X_n { \p W \over  \p P}
+ Q_n { \p W \over \p K} \right]
&(5.9) \cr }
$$
Here, $\tilde D$ is a modified phase space operator, appropriate
to the fact that the potential is $ Q V'(X)$, hence
$$
\tilde D = Q \sum_{n=1}^{\infty} { (-1)^n \over 2^{2n} } {1 \over (2n+1)!}
{d^{2n+2} V (X) \over dX^{2n+2} }
{ \p^{2n+1}  \over \p K^{2n+1} }
\eqno(5.10)
$$
This is the exact quantum dynamics of the deterministic quantum system
coupled to an environment. It is exactly decoherent in terms of
histories specifed by fixed values of $P,X,P_n,X_n$. It is
subject to the initial conditions that, in terms of the
original systems $A$, $B$, and their environments, the initial
state completely factors:
$$
W = W_A (p,x) W_B (k,y) W_{A\E} (p_n,q_n) W_{B \E} (\tilde p_n, \tilde
q_n)
\eqno(5.11)
$$
As in the previous section,
the auxiliary system $B$ is chosen to be in a minimum uncertainty
state. The environments of $A$ and $B$ are chosen to be
in thermal states, but with $T_A >> T_B$.

To compare with the standard quantum theory results (5.3) and (5.4),
we integrate out the variables, $K,Q,K_n,Q_n$. Tracing the
Wigner equation (to derive (5.4) from
(5.3), for example) is usually a non-trivial operation
[\cite{HaT}]. However, the fact that we not tracing out
canonical pairs appears to make it essentially trivial,
and it is easily seen that
the resulting Wigner
function $\tilde W(X,P,X_n,P_n)$ obeys the evolution equation
$$
\eqalignno{
{ \p \tilde W \over \p t} =&
- { P \over m} { \p \tilde W \over \p X} + V'(X) { \p \tilde W \over \p P}
\cr
& + \sum_n \left[ - {P_n \over m_n} { \p \tilde W \over \p X_n}
+ m_n \om_n^2 X_n { \p \tilde W \over \p P_n}
+ c_n  X { \p \tilde W \over \p P_n}
+ c_n X_n { \p \tilde W \over  \p P}
\right]
&(5.12) \cr }
$$
The evolution equations (5.12) and (5.3) are the same, except for the
term $D W $ in Eq.(5.3) (where note that the analagous term in
(5.9) dropped out when $K$ was integrated over).
In the absence of the environmental
terms, the presence of $ DW$ would substantially modify the
dynamics in (5.3) in comparison to (5.12). However, as
stated, after tracing out the environment in Eq.(5.3)
to yield (5.4),
the diffusive effects induced in the evolution of $W$
make the contribution of this term negligible. Moreover,
tracing out the environment in the DQT of Eq.(5.12)
leads to an equation of the form (5.4) without the term
$DW$, and with the temperature $T_A$ replaced by $T_A + T_B$.
The two evolution equations are therefore approximately the same
for $T_B << T_A$. We may therefore say the following:
the dynamics described by Eqs.(5.3) and (5.12) will be essentially
identical with respect to coarse-grainings asking questions
only about the variables $X$ and $P$. (We have phrased the
statement in this way, in terms of (5.12) and (5.3), rather than
(5.4) since the former are exact equations whereas (5.4)
holds only in the Fokker-Planck limit).

Given identical dynamics, the comparison of the two
systems then reduces to comparison of the initial states.
In the SQT result, Eq.(5.4), the initial state
is the Wigner function $W_A (p,x)$. In the corresponding
DQT equation (Eq.(5.12) with environment traced out),
by contrast, the initial state is the
reduced Wigner function,
$$
\eqalignno{
\tilde W (P,X) =& \int dQ dK \ W (K,X,P,Q)
\cr
& = \int d Q dK \ W_A (p,x) W_B (k,y)
&(5.13) \cr}
$$
This is written most usefully by changing variables
from $X,P,Q,K$ to $X,P,y,k$, where, from (5.6), we have
$$
Q = \half X - y, \quad K = P + k, \quad p = P + k, \quad x = X -y
\eqno(5.14)
$$
It follows that
$$
\tilde W (P,X) = \int d y dk \ W_A (P+k,X-y) W_B (k,y)
\eqno(5.15)
$$
Since $W_B$ is a minimum uncertainty state, this is a Wigner
function smeared over an $\hbar$-size region of phase space,
as in Eq.(4.11) (and is positive). We are therefore now
comparing the smeared Wigner function $\tilde W(P,X)$ which solves
the environment-traced version of
(5.12) to the Wigner function of the SQT, $W_A (p,x)$. These will
generally be different, but as stated in Section 4, the environment
comes to the rescue -- under evolution according to an
equation of the form (4.4), thermal fluctuations rapidly
overtake the quantum ones, and the difference between
the smeared and unsmeared Wigner functions is negligible.

We therefore have an independent proof of the approximate
equivalence of SQT and the DQT under the conditions
of approximate decoherence.

\head{\bf 6. Exact Decoherence of Momenta in the Quantum
Brownian Motion Model of SQT}

We now produce an example of a situation in standard
quantum theory, which does in fact exhibit exact
decoherence, without having to resort to the DQT of the
previous sections. The example is histories of momenta in the
quantum Brownian motion model, for a free particle
in the Fokker-Planck limit. It is in some ways a curious
and pathological example, but it does not appear to have
been noticed before, and is perhaps of interest in relation
to the discussions of the previous sections.

We first consider the form of the decoherence
functional for a system-environment model with, for simplicity,
projectors at two moments of time. It is
$$
D (\a_1, \a_2 | \a_1', \a_2 ) = \Tr \left( P_{\a_2}
K_0^t \left[ P_{\a_1} \rho P_{\a_1'} \right] \right)
\eqno(6.1)
$$
Here, the environment has been traced out, so the
projectors and the trace refer to the system only. The
evolution operator $K_0^t$ refers to reduced system dynamics
described by the master equation
whose Wigner transform is (4.4), that is, its solution is
$$
\rho_t = K_0^t [\rho_0 ]
\eqno(6.2)
$$
It is also useful to introduce a backwards time evolution operator
$ \tilde K_0^t $, defined by
$$
\Tr \left( A K_0^t [ \rho_0 ] \right) =
\Tr \left( \tilde K_0^t [ A ] \rho_0 \right)
\eqno(6.3)
$$
(this is not the inverse of $K_0^t $ since the evolution
is not unitary).
In terms of it, the decoherence functional may be written,
$$
D (\a_1, \a_2 | \a_1', \a_2 ) = \Tr \left( \tilde K_0^t \left[ P_{\a_2}
\right] P_{\a_1} \rho P_{\a_1'}  \right)
\eqno(6.4)
$$
Backwards evolution may also be described by a master equation
whose Wigner transform
is similar to the usual one (4.4), but the unitary and
dissipative terms have the opposite sign (we consider
only the case $V(x) = 0 $ here).
The decoherence term produces the same
effect in either direction in time.

By way of a digression,
from (6.4) we can see why decoherence of position histories is produced
by essentially the same mechanism that diagonalizes
the density matrix:
the projector
$ P_{\a_2} $ starts out diagonal in $x$ and remains approximately
diagonal in $x$ under evolution by $\tilde K_0^t$, hence when
acted on by position projectors $P_{\a_1}$, $P_{\a_1'}$ it
gives approximate diagonality of the decoherence functional.

After these preliminaries, turn to the case in which the projectors
in (6.4) are onto ranges of momenta. We shall show that diagonality
in $p$ is exactly preserved by $\tilde K_0^t$, for the case
of the free particle coupled to an environment in the Fokker-Planck
limit. To see this, consider first the Wigner representation of
the master equation in this case. It is
$$
{ \partial W \over \partial t} = - { p \over m}
{ \partial W \over \partial x} + 2 \gamma {\partial
\over \partial p} ( p W ) + 2 m \gamma k T
{ \partial^2 W \over \partial p^2 }
\eqno(6.5)
$$
The important property of this equation is the now following: if
$W$ is a solution to this equation, with initial condition
$W_0$, then $ { \partial W / \partial x } $ is also a solution, with
initial condition $ {\partial W_0 / \partial x } $. Translated
back into density operator language, this means that if
$ \rho_t$ is a solution to the master equation with
initial condition $ \rho_0$, then $[\rho_t, \ph ]$
is also a solution with initial condition $ [\rho_0, \ph]$, so
$$
[\rho_t, \ph ] = K_0^t \left[ [ \rho_0, \ph ] \right]
\eqno(6.6)
$$
This may also be written
$$
[ K_0^t \left[ \rho_0 \right], \ph ] =K_0^t \left[ [ \rho_0, \ph ] \right]
\eqno(6.7)
$$
or better,
$$
e^{ i a \ph} K_0^t [ \rho_0 ] e^{- ia \ph}
= K_0^t \left[ e^{i a \ph} \rho_0 e^{- i a \ph} \right]
\eqno(6.8)
$$
for any real constant $a$.

Now suppose that $[\rho_0, \ph] = 0$, which is equivalent to the
statement that $\rho_0$ is diagonal in $p$. Then it follows that
$[\rho_t, \ph ] = 0$ for all $t$. This manes
that the evolution operator $\tilde K_0^t$ preserves diagonality in
momenta. It follows immediately from this that the decoherence
functional (6.4) with projectors onto momenta will be {\it exactly}
diagonal.

Eq.(6.8) shows that the exact decoherence of momenta
comes from a translational invariance visible
in the path integral representation of $K_0^t$
(essentially Eq.(3.3) without the projectors,
with zero potential, and in the Fokker-Planck limit):
it is invariant under $x \ria x+a $,
$ y \ria y+a $.
Furthermore
it is broken by the frequency renormalization term in (3.6),
but we have here assumed that the renormalized frequency
is set to zero, along with the potential. This is all
rather unnatural, and for this reason this property is
an unphysical feature perhaps only of pedagogical value.
It ultimately traces back to the conservation
of momentum of the entire system (as long as the system
environment coupling is of the form $(x-q_n)^2$ in
Eq.(3.1))

The equivalent Langevin description also gives some insight.
The momenta, in this description, obey the equation
$$
\dot p + \gamma p = \eta (t)
$$
where $ \eta (t)$ is the usual Gaussian white noise. The important
point is that this equation is first order, so $p_t$ is a
function of $p$, but not of $\dot p$, so we expect in the quantum
theory that $ [\ph_t, \ph ] =0$, and therefore their histories
will be exactly decoherent.

On the other hand, whilst the density matrix (and indeed any other
evolving operator) will remain exactly diagonal in momenta, the
distribution of momenta $ \rho(p,p)$ will generally spread. We
therefore have the perhaps surprising situation of a quantity which
suffers fluctuations but is still exactly decoherent. The free
particle without an environment is clearly exactly decoherent in
momentum. Furthermore the distribution of momentum does not spread
for the free particle. On coupling to an environment in such a way
that the total (system plus environment) momentum is conserved, one
might expect to get only approximate decoherence of the system
momentum, since system momentum alone is no longer exactly
conserved. The surprise is that in a certain regime of
this model (the Fokker-Planck limit), the decoherence of momentum
{\it remains exact}, the environment making its mark only on the
momentum fluctuations which now do spread. This emphasises the fact
that the evolution of $ [\ph, \rho] $ (which controls decoherence)
and the evolution of $\rho(p,p)$ or $(\Delta p)^2$ (which controls
fluctuations) can really be quite different.

As stated, this example is in many ways a curiosity, but
it illustrates some interesting points. And in
the hunt for theories which are exactly decoherent it is
surely worth noting the places in which it was already lying
under our noses!

\head{\bf 7. A General Approach?}

We now turn to the question of how the construction
described may be extended to quantum systems which
are not described by a single simple canonical pair
satisfying (1.7), but instead by a more complicated
algebra. Spin systems, for example, are not described
by (1.7). Whilst we do not have a comprehensive answer
to this, the following is an indication of how one
might proceed.

Suppose
we have a quantum theory described by a set of
operators $ A_k$, $k=1,2 \cdots $ obeying a closed
algebra, where $[A_k,A_j] \ne 0 $ in general.
(The case described so far has
$A_1 = p$, $A_2 = x$, $A_3 = 1$.)
The equations of motion are
$$
\dot A_k = i [ H, A_k ] = f_k(A_1, A_2, \cdots)
\eqno(7.1)
$$
for some Hamiltonian $H = H(A_1, A_2, \cdots)$,
and the above relation defines the function $f_k$.
Suppose we consider the decoherence functional
for histories specified by fixed values of $A_k$.
Since $A_k$ at different times will generally not
commute, the histories will generally not be decoherent.

Now consider a second theory described by a set of
commuting operators $B_k$, with canonical momenta
$P_k$. Suppose that at the classical
level, they have the Poisson bracket relations,
$$
\{ B_k, B_j \} = 0, \quad \{ B_k, P_j \} = \delta_{kj},
\quad \{ P_k, P_j \} = 0
\eqno(7.2)
$$
Now define the Hamiltonian to be
$$
{\cal H} = \sum_k \ P_k \ f_k (B_1, B_2 \cdots )
\eqno(7.3)
$$
where $f_k$ is the function defined in (7.1). Then the
classical equations of motion for $B_k$ are
$$
\dot B_k = \{ B_k, {\cal H} \} = f_k (B_1, B_2, \cdots )
\eqno(7.4)
$$
On quantization (and with attention to operator
ordering), we thus obtain a set of commuting operators
$B_k$ which obey the same equations of motion
as the original set of operators $A_k$. This means that
histories of fixed $B_k$ will be exactly decoherent.
Furthermore, in the expression for the probabilities
for histories (1.1), the probabilities for histories
of $A_k$ and $B_k$ will be almost the same
function of the operators, differing in the form
of the initial state, and in the fact that the
trace in the case of the $B_k$ operators is over
a Hilbert space twice as large. Of course, these differences
may be substantial so this does not prove anything in
terms of the closeness of the two theories, but the above
shows that the question of the dynamics is straightforward.
A more
detailed description of the relationship between $A_k$
and $B_k$ is required for further analysis, and this is
perhaps best carried out with specific examples.
This will be pursued elsewhere.

\head{\bf 8. Summary and Discussion}

\subhead{\bf 8(A). Summary}

We have shown in a variety of ways that approximate decoherence of
histories of a system with canonical pair $p,x$ may be turned into
exact decoherence by doubling the Hilbert space and switching to the
classically equivalent variables $P = p-k$, $X = x+y$, where the
auxiliary variables $k,y$ are in a minimum uncertainty state. Any
non-decoherent set of histories may be made decoherent in this way,
but the point is that the change in the probabilities (or the Wigner
function) is small for histories which are already approximately
decoherent. The role of the environment in this scheme is that, by
giving the original system thermal fluctuations, it provides a kind
of ``smoke screen'' rendering the shift from $p,x$ to $P,X$
undetectable.

\subhead{\bf 8(B). An Alternative Approach to Emergent
Classicality?}

The approach described here might be regarded as giving an
alternative approach to emergent classicality.  Standard
demonstrations of approximate classicality involve comparing the
predictions of classical and quantum mechanics in a given
situation. Although this comparison is often clear intuitively,
at a more fundamental level the issue is perhaps clouded by the
fact that classical and quantum mechanics are theories of
different types: how can one measure the ``distance'' between
them? Here, however, in considering deterministic quantum theories
we are essentially writing down a quantum theory whose predictions
are exactly the same as a given classical theory.  To check for
emergent classicality we then compare standard quantum theory with
the deterministic quantum theory. Since the theories are the same
type of thing -- quantum theories -- it is clearer how they may be
compared. One may compare the density operators predicted by the
two theories, for example.

Although this conceptual advantage is admittedly minimal, there
could also be a practical advantage. The decoherence functional is
in general rather complicated to calculate, in comparison to
Wigner functions and density operators, to a degree that presents
problems in some areas of interest (such as the study of histories
of hydrodynamic variables [\cite{hydro}]). The results of this
paper suggest that a test for approximate decoherence of histories
consists quite simply of comparing the Wigner functions (or
density operators) of standard quantum theory and a suitably
chosen deterministic quantum theory.

\subhead{\bf 8(C). Other Approaches to Approximate Decoherence}

There are undoubtedly many other ways of investigating the
connection between approximate and exact decoherence, and it would
certainly be of interest to explore these. Here, we have adopted the
device of doubling the set of dynamical variables, and employed a
fundamentally different action. It would be of particular interest
to see whether one could avoid this in a simple way. For example,
the commuting position and momentum operators of von Neumann,
described in the Introduction, appear to hold the possibility of
moving from non-commuting to commuting operators without having to
change the underlying dynamics or the number of dynamical variables.

One of the difficulties of the present scheme is that the
Hamiltonian of the auxiliary system has wrong sign, leading to the
possibility of negative energies, although it is not clear that this
undesirable feature must arise in this context. For example, exactly
conserved quantities are exactly decoherent, and also the model of
Section 6 gave exact decoherence without negative energies.

An alternative scheme, similar to the present one, which avoids
negative energies is to add an identical auxiliary system with the
correct sign for the Hamiltonian, but work with complex
canonical variable [\cite{Ish}]. So we define
$$
\hat X = \hat x+ i \hat y, \quad \hat P = \hat p + i \hat k
\eqno(8.1)
$$
which clearly satisfy $[\Xh,\Ph]=0$.
The total Hamiltonian for a linear system is then
$$
H = { 1 \over 2 m} \Ph^{\dag} \Ph + \half m \om^2 \Xh^{\dag} \Xh
\eqno(8.2)
$$
which is positive. The difficulty with this approach
(although not obviously unsurmountable) is that now one is
faced with the issue of interpreting position
and momentum operators with an imaginary part.

These examples and their problems cause one to wonder whether all
attempts to distort approximate decoherence into exact decoherence
in a reasonably general way (i.e., not just for special initial
states) will encounter features which are difficult to accept.  We
might expect difficulties because we are in a sense trying to
rewrite quantum theory in essentially classical terms, and this is
well-known to lead to problems. Another example of this is the
Wigner function representation, which gives a deterministic
evolution equation close to the classical one for a phase space
distribution function, but it is not always positive so cannot be
directly interpreted as a true probability distribution.  Then
there is the Bohm theory approach to quantum theory which gives a
direct interpretation of the wave function  in terms of
trajectories, but is explicitly non-local. The extent to which
quantum theory cannot be interpreted in classical terms is
elegantly summarized in the Bell inequalities (and other related
results). This raises the question of whether inequalities of the
Bell type have something to say about the degree to which
decoherence may be made exact. These and other issues will be
explored elsewhere.

\head{\bf Acknowledgements}

I am grateful to Massimo Blasone, Chris Isham and Ray Rivers
for useful discussions.

\references

\def\pr{{\sl Phys.Rev.}}
\def\jmp{{\sl J.Math.Phys.}}

\refis{AnHa} C.Anastopoulos and J.J.Halliwell,
{\sl Phys.Rev.} {\bf D51}, 6870 (1995).

\refis{Bla} M.Blasone, P.Jizba and G.Vitiello,
hep-th/0007138.

\refis{CaL} A.Caldeira and A.Leggett, {\sl Physica} {\bf 121A}, 587 (1983).

\refis{DoH} H.F.Dowker and J.J.Halliwell, {\sl Phys. Rev.}
{\bf D46}, 1580 (1992).

\refis{DoK} The status of approximate decoherence (or
consistency) in the formalism of the decoherent histories approach
has been discussed by H.F.Dowker and A.Kent, {\sl J.Stat.Phys.}
{\bf 82}, 1575 (1996). They conjecture that an approximately {\it
consistent} set of histories are close to an exactly consistent
set. R.Omn\`es (private communication) has pointed out in many
models the real part of the decoherence functional is typically
highly oscillatory as a function of the times of the projections,
hence if it is close to zero, it can be made exactly zero by a
small change in the times. Here, however, we are interested in
decoherence in the more robust sense that it holds for a variety
of initial states and comes about because of an identifiable
physical mechanism, and nothing so far has been proved about this.

\refis{FeV} R.P.Feynman and F.L.Vernon,
{\sl Ann. Phys.} {\bf 24}, 118 (1963).

\refis{GeH1} M.Gell-Mann and J.B.Hartle, in {\it Complexity, Entropy
and the Physics of Information, SFI Studies in the Sciences of Complexity},
Vol. VIII, W. Zurek (ed.) (Addison Wesley, Reading, 1990); and in
{\it Proceedings of the Third International Symposium on the Foundations of
Quantum Mechanics in the Light of New Technology}, S. Kobayashi, H. Ezawa,
Y. Murayama and S. Nomura (eds.) (Physical Society of Japan, Tokyo, 1990).

\refis{GeH2} M.Gell-Mann and J.B.Hartle, {\sl Phys.Rev.} {\bf D47},
3345 (1993).


\refis{Gri} R.B.Griffiths, {\sl J.Stat.Phys.} {\bf 36}, 219
(1984); {\sl Phys.Rev.Lett.} {\bf 70}, 2201 (1993); {\sl
Phys.Rev.} {\bf A54}, 2759 (1996); {\bf A57}, 1604 (1998).

\refis{Hal1} J.J.Halliwell, in {\it Fundamental Problems in Quantum
Theory},  edited by D.Greenberger and A.Zeilinger, Annals of the New
York Academy of Sciences, Vol 775, 726 (1994),
gr-qc/9407040.

\refis{Hal5} J.J.Halliwell, {\sl Phys.Rev.} {\bf D60}, 105031
(1999).

\refis{Hal6} J.J.Halliwell,
{\sl Phys.Rev.} {\bf 46}, 1610 (1992).

\refis{HaT} J.J.Halliwell and T.Yu,
{\sl Phys.Rev.} {\bf 53}, 2012 (1996).

\refis{HaZ} J.J.Halliwell and A.Zoupas, {\sl Phys.Rev.}
{\bf D52}, 7294 (1995); {\bf D55}, 4697 (1997);
A.Anderson and J.J.Halliwell,
{\sl Phys.Rev.} {\bf D48}, 2753 (1993).


\refis{Har2} J.B.Hartle, in {\it Proceedings of the 1992 Les Houches Summer
School, Gravitation et Quantifications}, edited by B.Julia and J.Zinn-Justin
(Elsevier Science B.V., 1995), gr-qc/9304006.

\refis{Har6} J.B.Hartle, in, {\it Proceedings of
the Cornelius Lanczos International Centenary Confererence},
edited by J.D.Brown, M.T.Chu, D.C.Ellison and R.J.Plemmons
(SIAM, Philadelphia, 1994),
gr-qc/9404017 (1994).


\refis{HLM} J. B. Hartle, R. Laflamme and D. Marolf,
\pr {\bf D51}, 7007 (1995).

\refis{Hoo} G.'t Hooft, e-print hep-th/0003005,
quant-ph/9612018,
{\sl Class.Quant.Grav.} {\bf 16}, 3263 (1999).

\refis{Hu} B.L.Hu and Y.Zhang,
{\sl Int.J.Mod.Phys.} {\bf A10}, 453 (1995);
{\sl Mod.Phys.Lett.} {\bf A8}, 3575 (1993).

\refis{HPZ} B.L.Hu, J.Paz and Y.Zhang, {\sl Phys.Rev.D} {\bf 45},
2843(1992); {\bf 47}, 1576(1993).

\refis{HSZ} S.Habib, K.Shizume and W.Zurek,
{\sl Phys.Rev.Lett.} {\bf 80}, 4361 (1998).

\refis{hydro} J.J.Halliwell, {\sl Phys.Rev.Lett} {\bf 83}, 2481 (1999);
{\sl Phys.Rev.} {\bf D58}, 105015 (1998);
T.Brun and J.J.Halliwell,
{\sl Phys.Rev.} {\bf 54}, 2899 (1996);
T.Brun and J.B.Hartle, {\sl Phys.Rev.} {\bf D60}, 123503 (1999);
C.Anastopoulos, gr-qc/9805074.

\refis{Ish} C.Isham (private communication).

\refis{JoZ} E.Joos and H.D.Zeh, {\sl Z.Phys.} {\bf B59}, 223 (1985).

\refis{LowT} J.R.Anglin, J.P.Paz and W.H. Zurek, quant-ph/9611045.

\refis{McE} J.N.McElwaine, {\sl Phys.Rev.} {\bf A53}, 2021 (1996).

\refis{Neu} J. von Neumann, {\it Matematische Grundlagen der
Quantenmechanik} (Springer, Berlin, 1932). The relevant section is
reprinted in {\it Quantum Theory and Measurement}, edited by
J.Wheeler and W.Zurek (Princeton University Press, Princeton, New
Jersey, 1983)

\refis{Omn} R. Omn\`es, {\sl J.Stat.Phys.} {\bf 53}, 893 (1988);
{\bf 53}, 933 (1988); {\bf 53}, 957 (1988); {\bf 57}, 357 (1989);
{\sl Ann.Phys.} {\bf 201}, 354 (1990); {\sl Rev.Mod.Phys.} {\bf
64}, 339 (1992).

\refis{PaZ} W.H.Zurek and J.P.Paz, quant-ph/9502029.

\refis{QBM} G.S.Agarwal, \pr {\bf A3}, 828 (1971); \pr {\bf A4},
739 (1971); H.Dekker, \pr {\bf A16}, 2116 (1977); {\sl Phys.Rep.}
{\bf 80}, 1 (1991); G.W.Ford, M.Kac and P.Mazur, \jmp {\bf 6}, 504
(1965);H.Grabert, P.Schramm, G-L. Ingold, {\sl Phys.Rep.} {\bf
168}, 115 (1988); V.Hakim and V.Ambegaokar, \pr {\bf A32}, 423
(1985); J.Schwinger, \jmp {\bf 2}, 407 (1961); I.R.Senitzky, \pr
{\bf 119}, 670 (1960).

\refis{Wig} E. P. Wigner, {\sl Phys.Rev} {\bf 40}, 749 (1932).
For extensive reviews of the Wigner function and related
functions, see N. L. Balazs and B. K. Jennings, {\sl Phys.Rep.}
{\bf 104}, 347 (1984), and M. Hillery, R. F. O'Connell, M. O.
Scully and E. P. Wigner, {\sl Phys.Rep.} {\bf 106}, 121 (1984).

\refis{Zur1} J.P.Paz and W.H.Zurek, \pr {\bf D48},
2728 (1993); W.Zurek, in {\it Physical Origins of Time Asymmetry},
edited by  J.J.Halliwell, J.Perez-Mercader and W.Zurek (Cambridge
University Press, Cambridge, 1994);
in {\it Quantum Optics,
Experimental Gravitation and Measurement Theory}
(page 87), edited by P.Meystre and M.O.Scully
(NATO ASI Series, Plenum, New York, 1983);
{\sl Phil.Trans.R.Soc.Lond.A} {\bf 356}, 1793 (1998).
Zurek's contributions
to the subject are extensive and the above is only a
suggestive selection.




\endreferences

\end